# Accelerating Structure-Property Relationship Discovery with Multimodal Machine Learning and Self-Driving Microscopy

*Jiawei Gong[1,2], Danqing Ma[3], Ralph Bulanadi[1], Robert Moore[4], Rama Vasudevan[1], Lianfeng Zhao[3], Yongtao Liu[1*]*

[1] *Center for Nanophase Materials Sciences, Oak Ridge National Laboratory, Oak Ridge, TN, 37830, USA*

[2] *Department of Mechanical Engineering, Pennsylvania State University, Erie, PA, 16563, USA*

[3] *Holcombe Department of Electrical and Computer Engineering, Clemson University, Clemson, SC, 29634, USA*

[4] *Materials Science and Technology Division, Oak Ridge National Laboratory, Oak Ridge, TN, 37830, USA*

* Corresponding Author Email: liuy3@ornl.gov

**Abstract**

Microscopy combined with local spectroscopy is widely used to correlate nanoscale structure with functional properties in materials, but conventional measurements rely heavily on human-selected sampling locations and predefined targets, limiting dataset diversity and the potential for discovery. Here, we present a framework that integrates autonomous microscopy with a dual-novelty deep kernel learning (DN-DKL) for adaptive data acquisition and a dual variational autoencoder (VAE) for representation learning. DN-DKL actively guides the microscopy toward structurally and spectroscopically novel regions, enabling efficient collection of large spectral datasets. Dual-VAE embeds local structure and spectroscopic responses into a shared latent manifold that serves as a structure-property relationship map. We applied this framework for the investigation of halide perovskite films using conductive atomic force microscopy. The results reveal distinct hysteresis behaviors that are linked to specific nanoscale structural motifs, including grain boundary junction points that show hysteresis under different bias conditions and asymmetric grain boundaries that suppress charge transport. This framework establishes a general strategy that leverages the complementary strengths of self-driving microscopy, machine learning, and human expertise to accelerate scientific discovery in functional materials.

**Keywords**

Self-Driving Laboratory; Novelty Discovery; Representation Learning; Hybrid Perovskites; Atomic Force Microscopy

## Introduction

Microscopy has been a cornerstone in materials science and nanoscience, enabling researchers to investigate relationships between micro- and nanoscale structures and the resulting properties of functional materials [1–4]. Microscopy image analysis combined with local spectroscopy is widely used to establish structure-property relationships in materials. For example, in ferroelectric materials, piezoresponse force microscopy imaging reveals the domain structure, while local piezoresponse spectroscopy analyzes polarization switching behavior such as nucleation voltage and coercive field at specific domains, directly linking domain structures to polarization switching dynamics [5–7]. In nanoparticle samples, high-resolution scanning transmission electron microscopy images resolve atomic-scale defects, whereas spatially resolved electron energy-loss spectroscopy measures local band structure or carrier concentration, enabling the correlation between defect density and electronic transport properties [8–11]. In two-dimensional materials [12, 13], scanning probe microscopy (SPM) images lattice structure, edges, and wrinkles [14], while local conductive spectroscopy probes electronic density of states or excitonic behavior, directly connecting nanoscale structural features to local electronic properties.

Traditional correlative image and spectroscopy measurements in microscopy have relied on researcher expertise to guide where and what to measure, which are constrained by limited sampling, susceptibility to subjective bias, and the growing complexity of modern materials research. Recent advances in self-driving (or automated) microscopy offer a promising pathway toward more systematic exploration of structure-property relationships [15–21]. Automated microscopes can operate continuously, performing vast experiments with consistency that exceeds human capabilities, generating extensive datasets that would be prohibitively time-consuming to acquire manually [22, 23]. Furthermore, machine learning (ML) algorithms, such as supervised ML and active learning [24–29], can be incorporated into automated microscopy workflows to adapt data acquisition in real time, enabling purpose-driven data collection. Supervised ML can recognize and classify predefined structural features within microscopy images, enabling targeted selection of regions for subsequent measurements [30–33]. Supervised ML strategies are effective when the relevant structures and their significance are well understood, where it can guide the microscope toward structures deemed relevant based on prior knowledge and thereby streamlining data acquisition and improving efficiency. Active learning can guide experimental decisions by iteratively selecting new measurements that optimize a predefined property metric, such as maximizing conductivity, minimizing coercive field, or achieving a specific functional response. This approach is effective for directed optimization, but its reliance on predefined objectives inherently limits the scope of discovery. When focusing on predefined targets, active learning-based workflows may overlook unexpected phenomena or emergent behaviors.

In this work, we develop an integrated framework that combines the efficiency of automated microscopy for data acquisition, the capability of ML to disentangle complex, high-dimensional datasets, and the interpretive expertise of human scientists for structure-property relationship discovery. The framework incorporates a novelty-driven active learning strategy to accelerate the acquisition of diverse and informative measurements. We then employ a dual variational autoencoder (VAE) approach to jointly analyze structural information from microscopy images and local spectroscopic responses for systematic exploration of structure-spectroscopy relationships. Finally, domain experts interpret the learned correlations to extract physical insights and guide scientific understanding. We demonstrate the effectiveness of this approach by applying it to hybrid-perovskite thin-film materials, where the relationship between grain structure and current-voltage (IV) behavior plays a critical role in optoelectronic performance. This integrated workflow resolves distinct nanoscale electrical regimes associated with grain interiors, grain-boundary grooves, and triple-junction nano-traps, revealing that hysteresis and charge transport are governed more strongly by localized geometric trap structures than by average grain size alone.

## Results and Discussions

### *Multimodal MLs for Structure-Property Relationship Discovery*

A key challenge in discovering structure-property relationships lies in the intrinsic complexity of how material structure gives rise to functional behavior. This complexity arises because diverse structural features are often simultaneously coupled to multiple functional properties, making comprehensive exploration difficult. Functional materials usually exhibit heterogeneity across multiple structural dimensions at once. For example, at any given location, a material may be characterized by a specific grain size [34], grain boundary character [35], crystal facet orientation [36], domain configuration [37], and defect density. These features vary spatially and interact with one another, producing a heterogeneous microstructural landscape in which each local region possesses a distinct structural identity and local environment. In the meantime, a single material often exhibits multiple functional responses simultaneously, many of which can be encoded within the same local spectroscopic measurement and can be either dependent or interdependent. For instance, a polarization hysteresis loop can encode information about polarization magnitude, coercive field, and nucleation voltage, while a current-voltage (IV) curve may reflect conductivity, turn-on voltage, saturation current, and hysteresis behavior. Each of these properties may be governed by different structural factors or by complex interactions among multiple structural features. Consequently, structure-property relationships are rarely one-to-one; instead, they are inherently many-to-many (as shown in Figure 1) with intertwined dependencies between structural characteristics and functional responses.

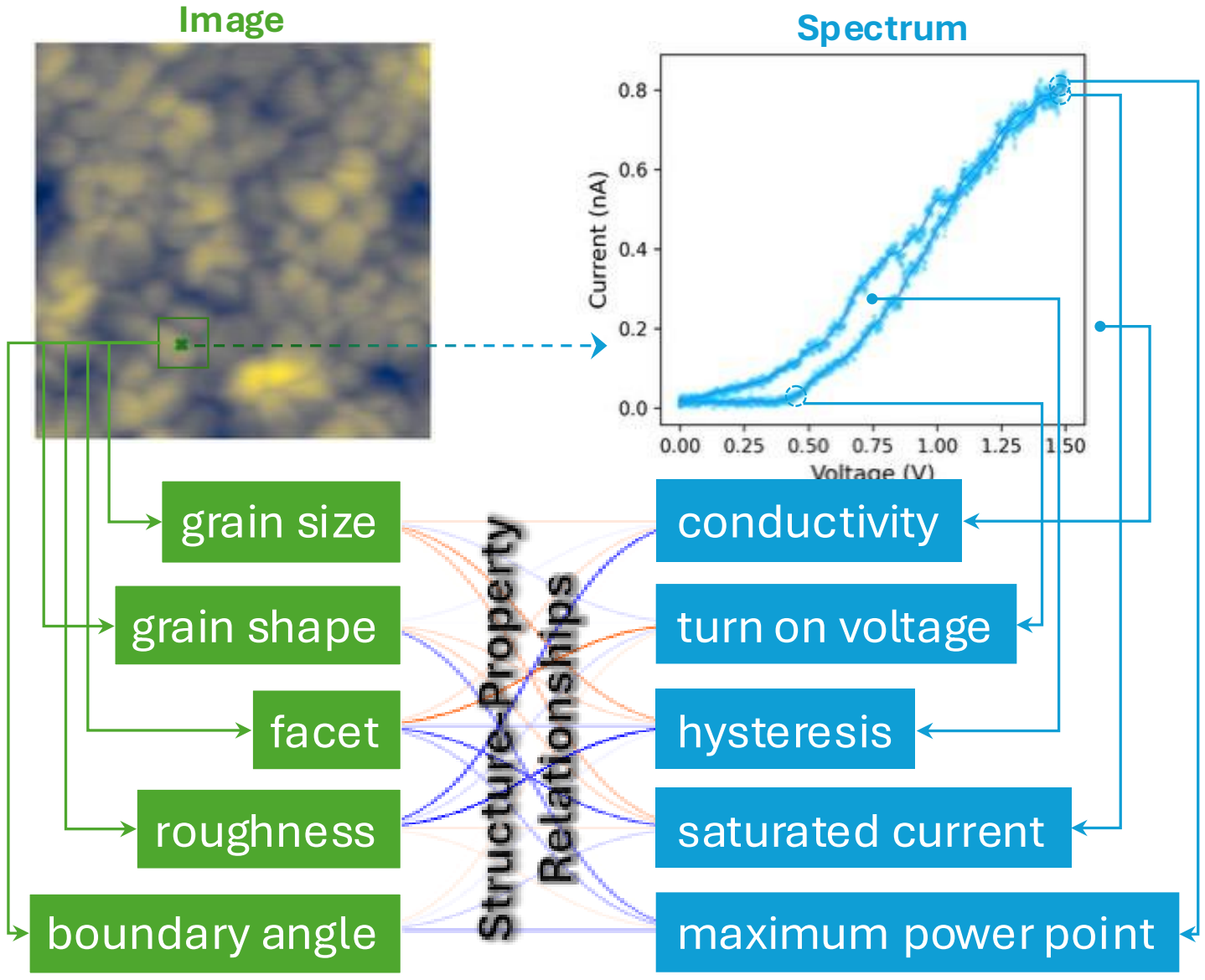

**Figure 1.** The complex, intertwined dependencies between structural characteristics and functional responses give rise to many-to-many structure-property relationships. Microscopy, combining high-resolution imaging with local spectroscopy measurement, is a powerful tool for probing these relationships at the micro- and nanoscale.

These considerations impose several key requirements on a framework aimed at discovering structure-property relationships. First, the framework must support the acquisition of structurally and functionally diverse datasets that adequately sample the heterogeneity of the material. Second, it must provide mechanisms to disentangle the complexity of the many-to-many structure-property mapping while minimizing information loss in both structural and property representations. To meet these requirements, we develop a framework (Figure 2a) that integrates automated microscopy with both a novelty discovery algorithm for accelerated data acquisition, and representation learning using a variational autoencoder (VAE) for complex correlation analysis; the resulting outputs are then assessed and interpreted by human experts to derive physical insights.

We first develop an approach to acquire datasets that capture both structural diversity and rich spectroscopic information. To achieve this, we employ deep kernel learning (DKL) [38, 39], which has demonstrated strong capability in driving autonomous data acquisition in microscopy experiments [40, 41]. In prior implementations, DKL is typically guided by a predefined property descriptor, enabling efficient optimization of the targeted property. While such a property-descriptor-driven approach allows exploration over a larger parameter space than manual measurements, it tends to follow narrow trajectories guided by predefined and specific metrics. As a result, it often fails to sample the broader landscape of structural and spectral information that extend beyond the predefined property metrics.

To overcome this limitation, we introduce novelty-based driving metrics into the DKL framework (Figure 2b). Specifically, we define a novelty score for spectroscopic measurements that quantifies how distinct a newly acquired spectrum is relative to previously collected data, thereby prioritizing the discovery of previously unexplored spectral responses [42]. In addition, we also introduce a novelty score for structural images to refine the decision-making process in the DKL framework, prioritizing regions with structural features that are distinct from those already measured. Novelty scores for unmeasured structural images are incorporated into the DKL acquisition function via a gated active learning framework [43, 44] to jointly select the next measurement point. In this framework, the spectral novelty score is used directly as the training target for DKL, while the image-based novelty score is incorporated into the decision-making process to prioritize measurement at new structures. Together, this novelty-prioritizing strategy ensures the acquisition of both diverse spectroscopic data and diverse structural configurations, which are essential for establishing comprehensive structure-property relationships. Hereinafter, this approach will be referred to as dual-novelty-DKL (DN-DKL).

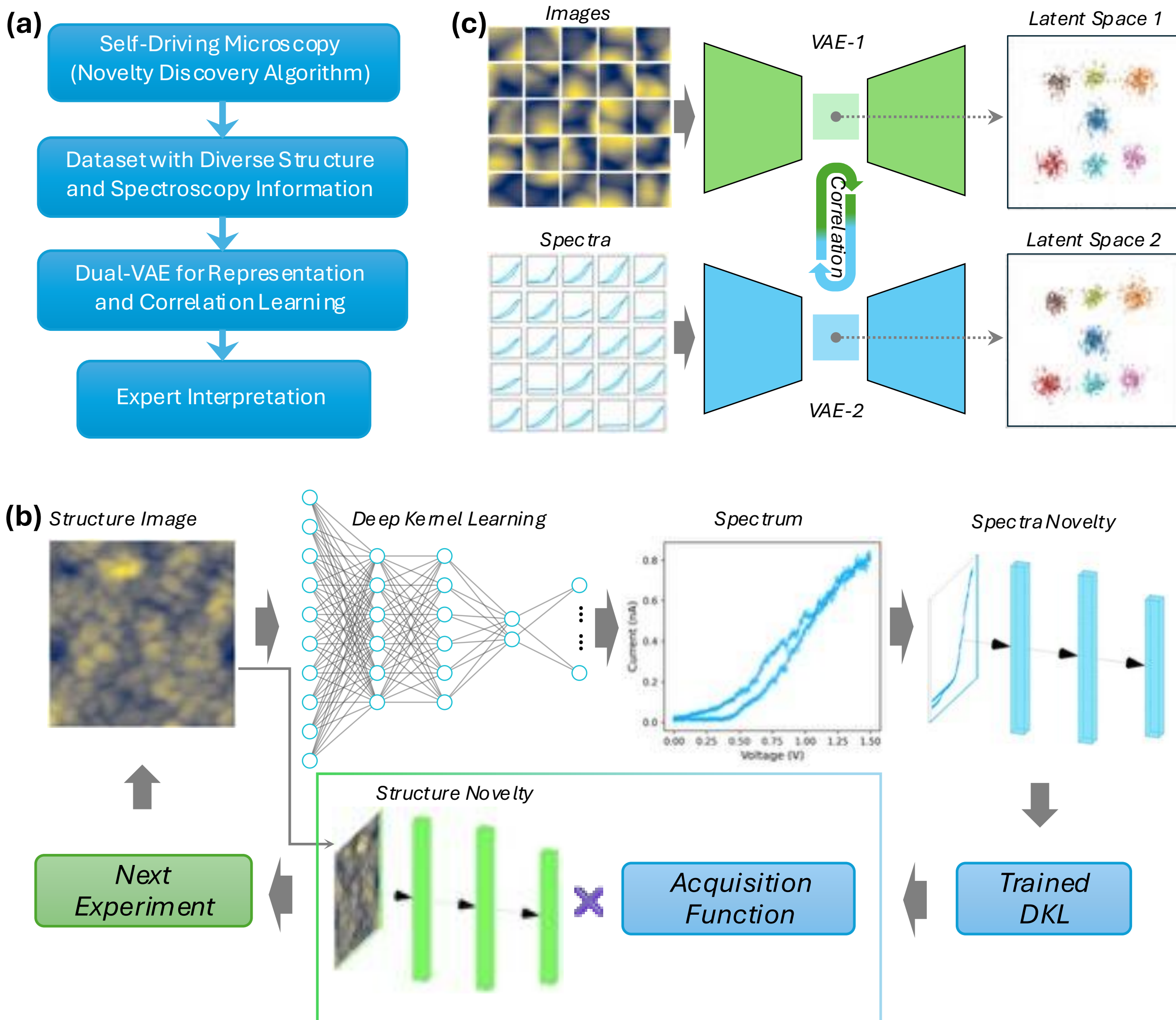

**Figure 2.** DN-DKL framework and Dual-VAE for structure-spectroscopy relationship investigation. (a) The workflow consisting of DN-DKL for data acquisition and Dual-VAE for disentangling the structure-property relationship. (b) DN-DKL workflow integrating spectral and structural novelty scores to guide measurement selection. Spectral novelty prioritizes acquisition of distinct spectra data, while structural novelty prioritizes measurements at structurally distinct locations. (c) Dual-VAE with parallel encoders for structural images and spectra, trained jointly to learn aligned latent representations and discover structure-property correlations.

After acquiring structure images and spectra data, we employ a dual variational autoencoder (Dual-VAE) [45] to learn correlations between structural images and spectroscopic data, as shown in Figure 2c. The Dual-VAE consists of two parallel VAEs that encode structural images and spectra, respectively. Each VAE follows a standard encoder-decoder design, compressing high-dimensional inputs into low-dimensional latent distributions and reconstructing the original data. The two VAEs are jointly trained with aligned latent representations. By jointly optimizing reconstruction and latent consistency losses, the dual-VAE learns a correlated structure-spectra manifold that integrates complementary information from both structural images and spectroscopic data. This approach enables disentangling of the structure-property relationships in complex materials systems.

*Application to Halide Perovskite Thin Films*

We applied this workflow to investigate halide perovskite thin films using conductive atomic force microscopy (cAFM). Halide perovskites have promising applications in solar cells, light-emitting diodes, and photodetectors due to their exceptional optoelectronic properties. These materials exhibit complex nanoscale behavior with spatially and temporally heterogeneous electrical responses. Local properties such as conductivity, band alignment, and recombination dynamics can vary significantly across a single film despite uniform macroscopic measurements. This heterogeneity is closely associated with nanoscale and microscale features such as grains, grain boundaries, surfaces, and interfaces, where defect distribution, chemical composition, and lattice strain are often nonuniform. These regions can either facilitate charge transport or act as sites of ion accumulation and nonradiative recombination, and their roles may evolve dynamically under bias. The cAFM is a powerful tool for investigating charge transport in perovskites at the nanoscale. In cAFM, topographic imaging captures the local structure of grains and grain boundaries, while localized IV measurements at selected positions reveal electrical transport responses. By correlating IV characteristics with structural features, cAFM enables investigation of the relationship between morphology and electrical response.

We investigated a mixed-cation halide perovskite thin film $Cs_{0.05}MA_{0.05}FA_{0.90}PbI_3$. The film was prepared from a precursor solution of formamidinium iodide (FAI), methylammonium iodide (MAI), cesium iodide (CsI), and $PbI_2$ dissolved in a DMF:DMSO mixed solvent (4:1 v/v). The films were deposited onto ITO substrates modified with a [2-(4-methoxyphenyl)ethyl]phosphonic acid (MeO-2PACz) self-assembled monolayer, serving as an efficient hole transport layer. Scanning electron microscopy (SEM) imaging (Figure 3a) reveals a compact, continuous morphology with dense grain packing, characteristic of high-quality polycrystalline films. Time-resolved photoluminescence (TRPL) measurement (Figure 3b) reveals the kinetics of carrier dynamics. By fitting the decay curve with a tri-exponential model [46, 47], we found that the process is dominated by a fast component ($\tau_1$ = 1.28 ns; ~59% amplitude), likely corresponding to rapid trap-assisted recombination at the interface or surface. This is accompanied by intermediate and slow components ($\tau_2$ = 6.81 ns and $\tau_3$ = 39.76 ns), yielding an average carrier lifetime ($\tau_{average}$) of 25.81 ns. The presence of the long-lived $\tau_3$ component indicates substantial bulk crystalline quality, allowing a significant portion of carriers to survive for collection. In addition, 2D PL lifetime mapping (Figure 3c) reveals spatial inhomogeneity in carrier lifetimes; these nanoscale variations underscore the necessity of detailed, spatially resolved structure-spectroscopy relationships investigated in this work.

To further validate the quality of the hybrid perovskite film, we fabricated perovskite solar cells with a standard architecture of ITO/MeO-2PACz/Perovskite/$C_{60}$/BCP/Ag (Figure 3d). The devices achieved a power conversion efficiency of 20.38%. Notably, this performance was obtained without additives or surface passivation layers, indicating that the measured characteristics primarily reflect the intrinsic behaviors of the perovskite material.

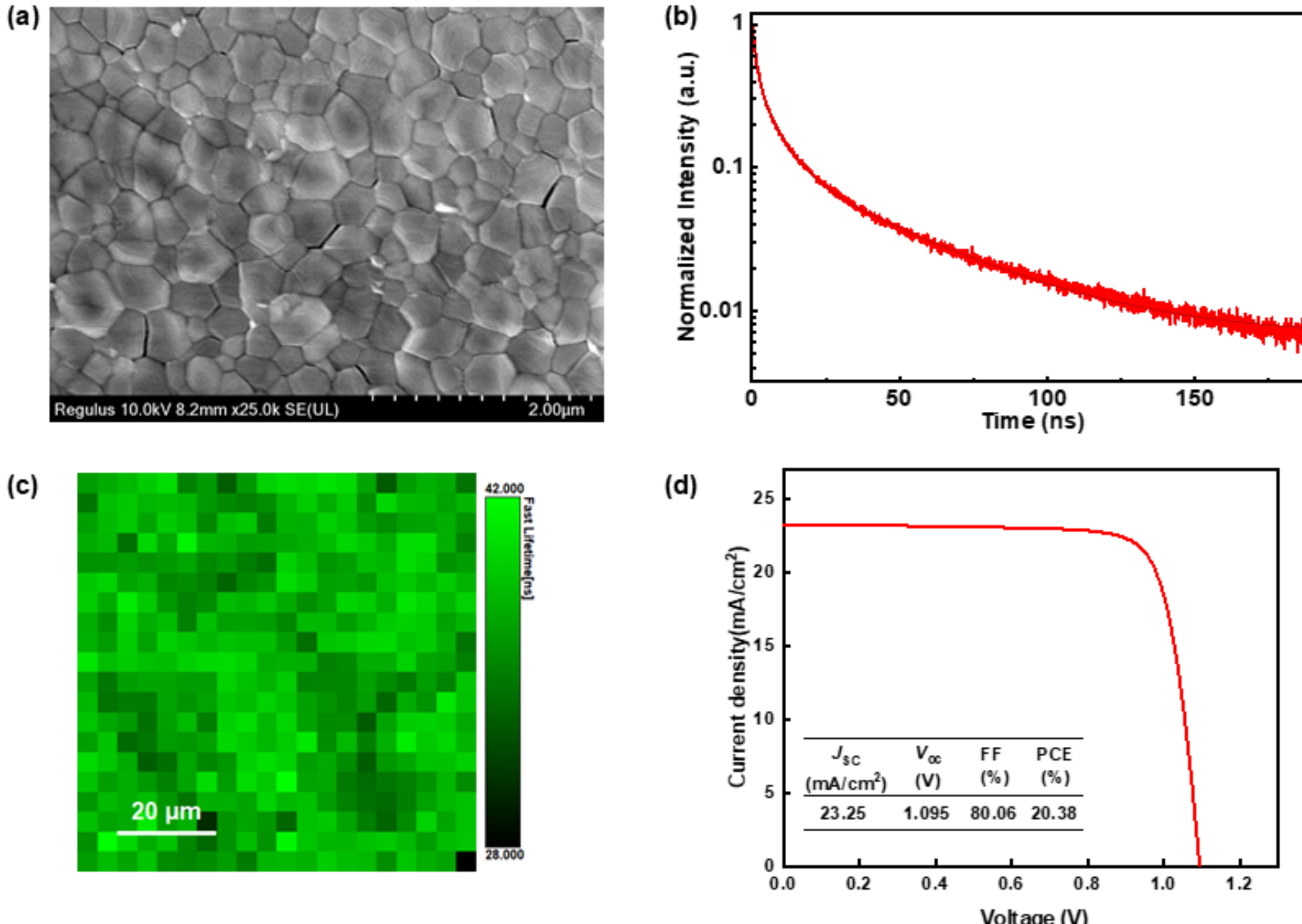

**Figure 3**. Characterization of the perovskite films and solar cell performance. (a) Top-view scanning electron microscopy (SEM) image of the perovskite film, displaying a dense and compact polycrystalline morphology. (b) Time-resolved photoluminescence (TRPL) decay kinetics of the perovskite film. (c) Spatially resolved PL lifetime mapping, highlighting nanoscale variations in carrier dynamics and film homogeneity. (d) Current density-voltage (*J-V*) curve of the perovskite solar cell measured under simulated AM 1.5G illumination.

*DN-DKL Autonomous cAFM Acquisition*

We applied DN-DKL to cAFM-IV data acquisition on the halide perovskite thin film. Conventional cAFM-IV measurements rely on manual operation, in which human operators position the AFM probe at selected locations to acquire IV data. This procedure inherently limits dataset size and diversity, making a comprehensive investigation of structure-spectroscopy relationships impractical. In contrast, DN-DKL enables autonomous IV measurements that actively seek diverse and previously unseen behaviors. We implemented DN-DKL in our autonomous AFM system through our AEcroscopy platform [15], which enables seamless deployment of various ML approach for automated and autonomous microscopy experiments. In the DKL setting, the model is trained using local topography image patches that capture micro- and nanoscale structural features, and IV curves that encode the electrical response. The dual-novelty strategy ensures that the system prioritizes acquisition in regions that are simultaneously structurally unfamiliar and spectroscopically distinct from previously measured locations. As a result, the DN-DKL-driven autonomous cAFM samples rare features and underrepresented behaviors, producing a large, information-rich dataset with broad morphological and electrical diversity for comprehensive analysis of structure-spectroscopy relationships.

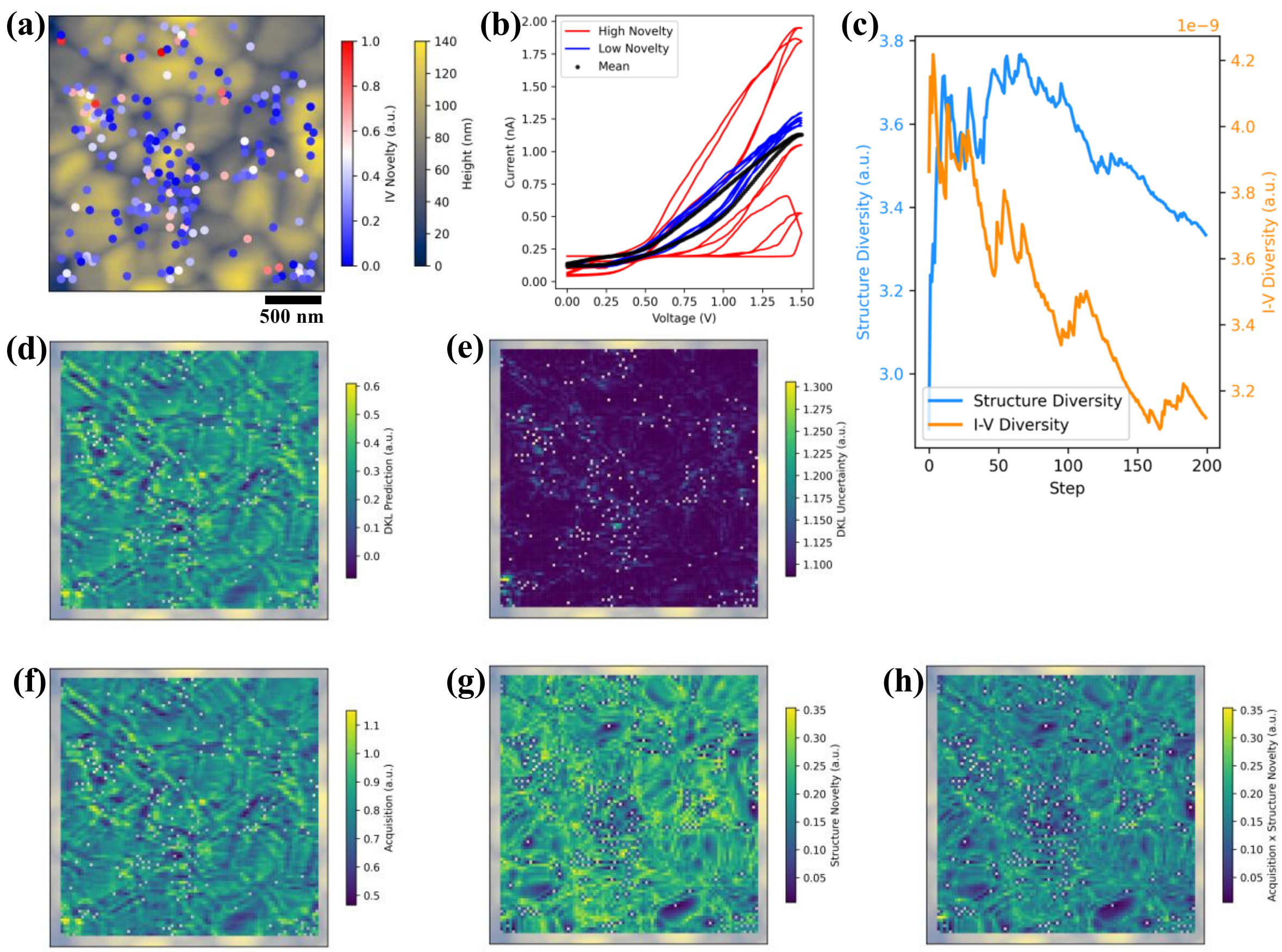

**Figure 4.** DN-DKL exploration of halide perovskite thin films. (a) Topography with overlaid IV measurement locations. (b) Representative high-novelty (top 5) and low-novelty (bottom 5) IV curves compared to the mean IV curve. (c) Structural and spectral diversity versus exploration step, showing initial rapid increase followed by gradual decrease with intermittent rises. (d–e) IV novelty prediction and uncertainty maps after 200 steps. (f–h) Acquisition function, image novelty score, and combined maps for unmeasured locations.

Figure 4 shows the DN-DKL experimental process and results. Figure 4a shows the topography image containing structural features such as grains and grain boundaries. Overlaid on the topography are the IV measurement locations selected sequentially by DN-DKL. Starting from 10 seed points, 200 DN-DKL exploration steps were performed, yielding a total of 210 IV measurements. Figure 4b compares representative high-novelty and low-novelty IV curves against the mean IV curve; we plotted the top 5 high-novelty and 5 low-novelty IV curves. The high-novelty IV curves exhibit distinct features and deviate significantly from the mean IV, in contrast, the low-novelty IV curves closely resemble both the mean and each other. This suggests that spectral-novelty-driven exploration effectively directs the measurement toward the discovery of IV characteristics distinct from previously sampled behavior.

Dataset diversity can be another metric for assessing the novelty-driven process. Figure 4c plots the diversity of both the structural patch set and the IV spectra set as a function of DN-DKL step.

Here, both structural and spectral diversity increase drastically during the initial steps, reflecting effective novelty-driven exploration. After approximately 20 steps, diversity gradually decreases; notably, in a finite experimental space, overall dataset diversity is expected to decrease after certain sampling steps. However, sporadic increases in both structural diversity and spectral diversity continue to occur, indicating that the novelty-driven approach continuously enhances diversity of acquired datasets. The steps at which structural diversity and spectral diversity increase do not always coincide, suggesting that the structural novelty score and spectral novelty score contribute at different stages and work synergistically to maximize dataset diversity.

Figure 4d–e shows the DN-DKL prediction and uncertainty maps after 200 experiments. The predicted IV novelty shows a correlation with grain structure (Figure 4a), suggesting a strong structure-spectroscopy relationship that will be analyzed further later. Figure 4f–h shows the DN-DKL acquisition function, image novelty score map, and the combined map of acquisition and image novelty for unmeasured locations, respectively. The image novelty score map clearly shows lower structural novelty around measured points (where the measured points are the white markers), indicating that the structural novelty effectively directs measurements toward new structural features rather than repetitively sampling similar structures. Together, these results demonstrate that DN-DKL effectively acquires datasets with diverse structural features and spectroscopic characteristics.

*Dual-VAE Structure-Property Relationship Analysis*

Figure 5 shows the cAFM topography-IV dataset of structural images and IV curves acquired by DN-DKL. Figure 5a is the topography overlaid with IV measurement locations, color-coded by sampling step. Figure 5b shows the structural patches from each measurement location. This image set reveals substantial structural diversity: some patches contain large grains, some capture grain boundaries or boundary junction points. Grain shapes, flatness, and boundary widths also vary across the dataset. Figure 5c shows the IV curves, which exhibit diverse characteristics including variations in current magnitude, current-voltage hysteresis behavior, and in some cases, negligible current response.

These results acquired by DN-DKL-driven autonomous cAFM capture both structural and electrical heterogeneity, enabling the investigation of structure-property relationship using Dual-VAE. Figure 5d shows the combined latent space distribution, providing a joint representation of structural features and IV responses. Notably, the combined latent space is not just independent dimensionality reduction of structural features and IV responses; it also learns a coupled representation of the co-variation between structural features and IV behaviors. Different regions of the latent space reveal different levels of structure-IV coupling. In the top and right regions, the image and IV samples with the same color value cluster closely together (as indicated by dashed boxes), indicating the Dual-VAE clearly captured their correlation. In contrast, the bottom-left region does not exhibit clear correspondence between structures and IV responses; here similar structural features can lead to stochastic IV responses, or similar IV responses correspond to heterogeneous structures. This behavior may originate from hidden variables not directly resolved in topography such as subsurface defects or compositional fluctuations. Because of the weak correlation, we would not draw a direct conclusion about the structure-property relationship from the bottom-left region in our further analysis. Nevertheless, the presence of this weak correlation can still be scientifically meaningful; it may point to missing or unobserved physical descriptors, or mechanisms governing IV behaviors that are not morphological structure dependent.

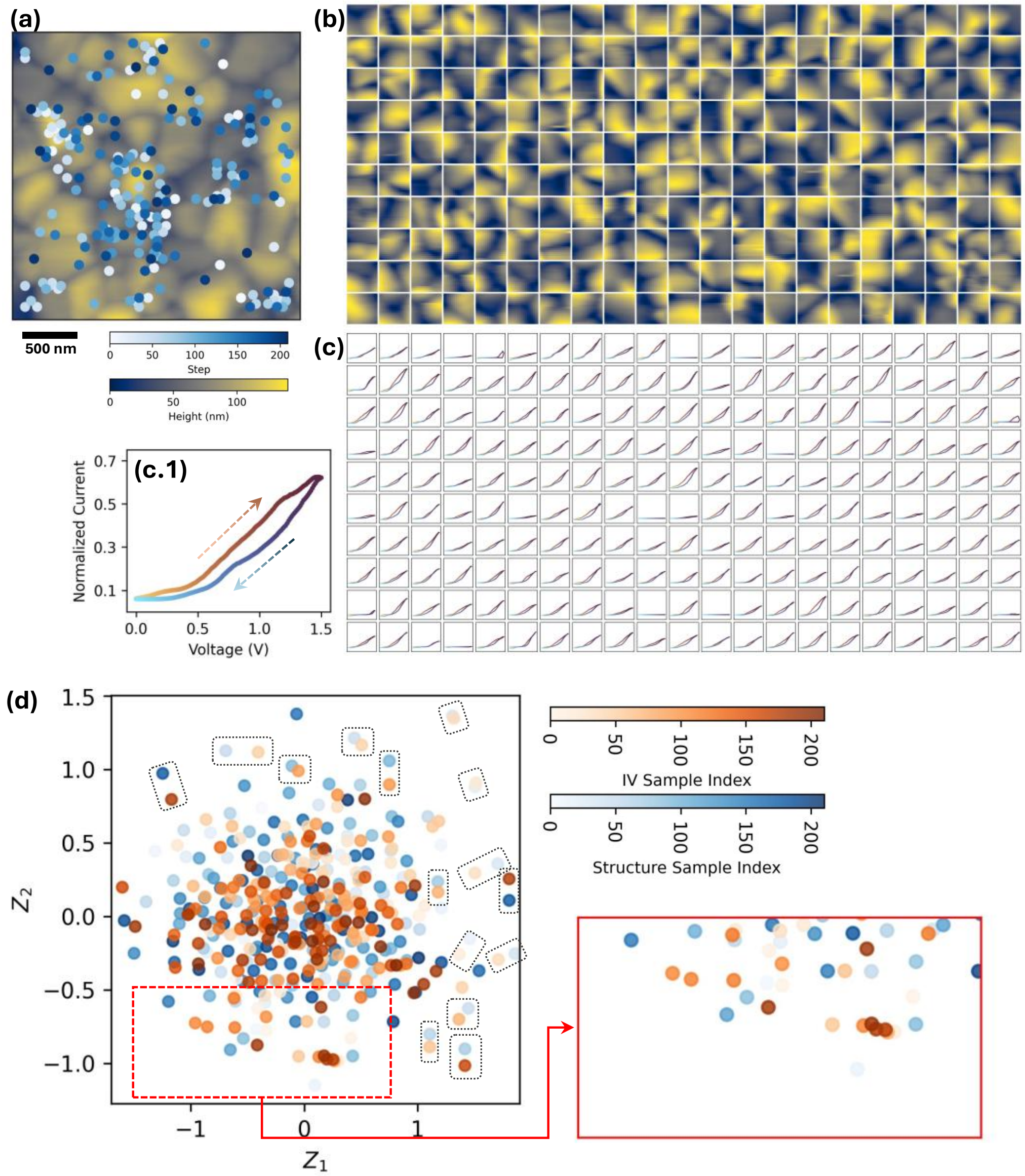

**Figure 5.** cAFM topography-IV Dataset. (a) Topography with color-coded measurement locations by sampling step. (b) Structural patches from the IV measurement locations showing diverse grain features and boundaries. (c) IV curves exhibiting varied current magnitudes, hysteresis, and conductivity, (c.1) indicates the voltage sweep direction of IV curves. (d) Dual-VAE latent space indicates the learned correlation of structural features and IV responses, where strong correlation between structure and IV is observed on top and right of the latent space (as exemplified by dashed boxes), but the bottom-left side indicates very weak correlation (as highlighted by red box).

By embedding structural patches and corresponding IV curves into correlated manifolds (shown in Figure 6a-b), the model organizes the dataset according to underlying physical correlation. As a result, nearby points in the manifold correspond to regions that exhibit similar IV behavior (Figure 6b) associated with similar nanoscale structural features (Figure 6a). Examination of the correlated manifolds (Figure 6a-b) reveals both known physical behavior and previously underexplored phenomena.

First, in the top-right region (highlighted by a red box) of the structural manifold, a continuous transition from grain interiors to grain boundaries is observed. To better resolve this trend, we refined the sampling density of the manifold from a 10x10 grid to a 30x30 grid and extracted the corresponding structural patches and IV curves from matched spatial locations, shown in Figure 6c. Notably, IV measurements were performed at the center of each structural patch. Moving from left to right in Figure 6c, the IV measurement locations (patch center point) gradually shift from grain interiors to grain boundaries. Correspondingly, the IV curves show a progressive increase in current-voltage hysteresis. This trend suggests that grain boundaries play a significant role in modulating electrical transport. The pronounced hysteresis is likely associated with trap states located in grain boundary grooves and/or junctions [48]. This observed behavior is consistent with the established understanding that grain boundaries in halide perovskites contribute strongly to hysteretic transport [49], thereby supporting the physical validity of the learned manifold representation.

In addition to the continuous transition, the entire IV manifold (Figure 6b) reveals three distinct hysteresis behaviors, representative IV curves labeled as "club", "heart", and "diamond". These are further illustrated in Figure 6d–f together with the corresponding structural features. The "club" behavior (Figure 6d) exhibits a hysteresis opening at intermediate bias (approximately 0.5-1.0 V); and the corresponding structure patches primarily indicate grain interiors. The "heart" behavior (Figure 6e) shows two hysteresis openings, one at low bias (~0-0.5 V) and the other at intermediate-to-high bias (~0.7-1.3 V); the corresponding structures are grain boundaries and triple-junction boundary points [50]. In contrast, the "diamond" behavior (Figure 6f) displays hysteresis at high bias (~1.3-1.5 V); and the associated structures are asymmetric grain boundaries, where one side grain forms a sharp edge and the other side forms a smoother, deeper boundary, as indicated in the schematic illustration. At these asymmetric boundaries, the turn-on voltage of current is much higher (~1.3 V) and the resulting current is strongly suppressed. This suggests the presence of a local energy barrier or band misalignment that inhibits charge transfer near these asymmetric boundaries [51]. By comparison, the "club" and "heart" behaviors exhibit lower turn-on voltages (~0.2 V) and higher resulting currents, indicating more efficient charge transport.

Overall, these results indicate that current-voltage hysteresis is strongly correlated with the presence and type of grain boundaries. Different boundary morphologies appear to produce distinct transport regimes, highlighting the importance of nanoscale interfacial structure in governing local electrical behavior. The physical mechanisms associated with the variations of grain boundary shapes, such as defect accumulation or local band bending, require further attention in future investigation, but the learned manifold provides a systematic landscape for future study.

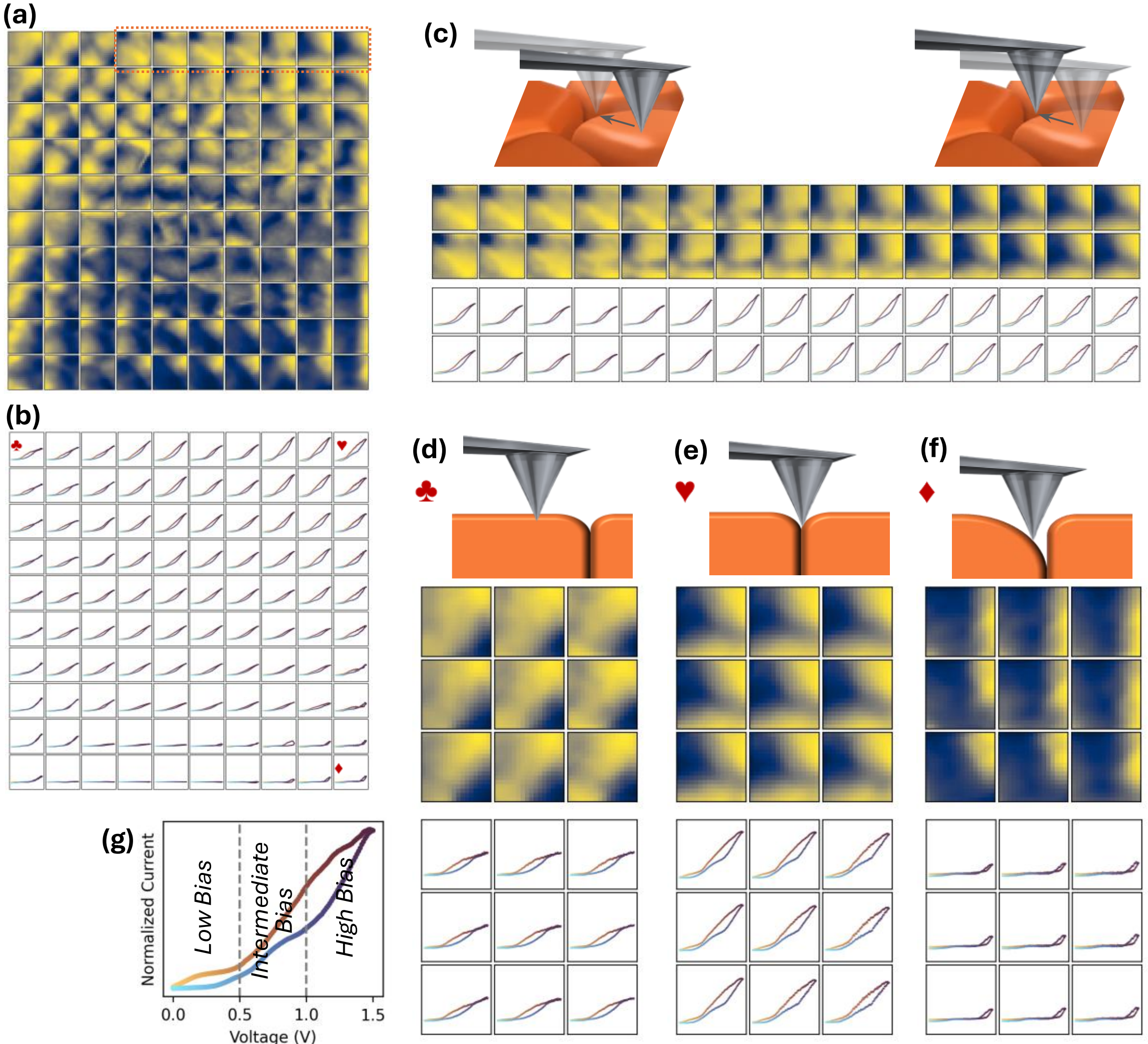

**Figure 6**. Correlated manifolds learned by the Dual-VAE linking structural features and IV curves. (a) Structural manifold, where each grid point represents a reconstructed structural feature corresponding to a latent coordinate. (b) IV manifold showing the reconstructed IV curves at the same latent coordinates as in (a), demonstrating a shared latent representation between structure and IV curves. (c) Higher-resolution sampling corresponds to the region marked with the red box in (a) of the manifold illustrating a continuous transition of measurement locations from grain interiors to grain boundaries; the corresponding IV curves show progressively increasing hysteresis. (d–f) Representative IV hysteresis behaviors identified from (b) marked as "club," "heart," and "diamond" with their corresponding structural features. The "club" behavior is associated with grain interiors, the "heart" behavior with grain boundaries and triple junctions, and the "diamond" behavior with asymmetric grain boundaries that exhibit high turn-on voltage and suppressed current. (g) indicates low, intermediate, and high bias conditions referred to in the discussion.

## Conclusion

In summary, we developed a framework for accelerating structure-property relationship discovery by integrating a discovery-driven workflow for autonomous data acquisition with correlative representation learning. The discovery-driven workflow, i.e., dual novelty deep kernel learning (DN-DKL), actively selected measurement locations based on both structural and spectroscopic novelty to acquire data with unexplored phenomena, which enables efficient acquisition of information-rich datasets. In addition, a dual variational autoencoder (Dual-VAE) embedded local structural features and spectroscopic responses into a shared latent space, providing a correlative map that links nanoscale structure to functional behavior. We applied this framework to autonomous conductive atomic force microscopy measurements of halide perovskite thin films and revealed previously under-sampled behaviors. The results revealed distinct current-voltage hysteresis responses associated with specific nanoscale motifs, including grain boundary junctions that exhibit bias-dependent hysteresis and asymmetric grain boundaries that suppress charge transport. These findings demonstrate how novelty-driven autonomous experiments, combined with representation learning, can accelerate scientific discovery. More broadly, this framework can be extended to incorporate other AI/ML-guided data acquisition strategies to enhance dataset diversity, and AI/ML-driven relationship discovery, hypothesis generation, and feature extraction. Future work could also extend the relationship analysis to multiple image channels and multiple spectra channels, allowing the ML model to identify intertwined relationship between multiple images and spectra that are challenging for human experts to recognize. This strategy establishes a general pathway toward self-driving laboratories and faster discovery in complex materials, by leveraging the efficiency of autonomous data acquisition, ML analysis, and human interpretation, which can be applied to a wide range of material systems and multimodal characterization techniques.

## Acknowledgements

This research was sponsored by the INTERSECT Initiative as part of the Laboratory Directed Research and Development Program of Oak Ridge National Laboratory, managed by UT-Battelle, LLC for the US Department of Energy under contract DE-AC05-00OR22725. Scanning probe microscopy experiments as well as data analysis research were supported by the Center for Nanophase Materials Sciences (CNMS), which is a US Department of Energy, Office of Science User Facility at Oak Ridge National Laboratory. This work was supported in part by the U.S. Department of Energy, Office of Science, Office of Workforce Development for Teachers and Scientists (WDTS) under the Visiting Faculty Program (VFP). L.Z. and D.M. acknowledge support from the National Science Foundation under Award Nos. DMR-2403802 and ECCS-2304364.

## Conflicts of Interest

The authors declare no conflicts of interest.

## Author contributions

Y.L. conceived the research. Y.L. designed and performed c-AFM experiments. D.M. and L.Z. grew and characterized the perovskite thin films and solar cells. J.G. performed data analysis. Y.L. and J.G. wrote the manuscript with support from all the co-authors.

## Data Availability

Data and code for this work are available in the manuscript, supplementary information, or at Repo: https://github.com/yongtaoliu/DNDKL_DVAE_PVSK.

## Experimental Methods

*Perovskite Precursor Preparation*

Formamidinium iodide (FAI) and methylammonium iodide (MAI) were purchased from Greatcell Materials. Lead(II) iodide ($PbI_2$, 99.99%) was purchased from TCI. Cesium iodide (CsI), N,N-dimethylformamide (DMF), dimethyl sulfoxide (DMSO), isopropanol (IPA), and chlorobenzene (CB) were obtained from Sigma-Aldrich. The self-assembled monolayer (SAM) material, [2-(4-methoxyphenyl)ethyl]phosphonic acid (MeO-2PACz), was purchased from Luminescence Technology Corp. (Lumtec). Fullerene (C60) and Bathocuproine (BCP) were purchased from Xian Polymer Light Technology Corp. All chemicals were used as received without further purification. The perovskite precursor solution (1.5 M) was prepared by dissolving 691.5 mg of $PbI_2$, 232.16 mg of FAI, 11.92 mg of MAI, and 19.49 mg of CsI in 1 mL of a mixed solvent of DMF and DMSO (4:1 volume ratio). The solution was stirred at room temperature until fully dissolved and filtered through a 0.45 μm PTFE filter prior to use.

*Device Fabrication*

Hole Transport Layer (HTL): The MeO-2PACz SAM solution was prepared by dissolving the powder in IPA at a concentration of 0.3 mg/mL. The SAM solution was spin-coated onto the substrate at 3000 rpm for 30 s with an acceleration of 1500 rpm/s.
Perovskite Active Layer: The perovskite films were deposited using a two-step spin-coating program: (1) 2000 rpm for 10 s (acceleration 200 rpm/s) and (2) 4000 rpm for 40 s (acceleration 1500 rpm/s). During the second step, 150 μL of CB antisolvent was dripped onto the center of the substrate at the 35th second from the start of the program. The films were then immediately annealed on a hotplate at 100 ℃ for 30 min.
Electron Transport Layer (ETL) and Electrode: The devices were completed by thermally evaporating C60 (30 nm), BCP (8.5 nm), and Ag (85 nm) in a thermal evaporator. The deposition rates were controlled at 0.3 Å/s for both C60 and BCP, and the Ag electrode was evaporated at a rate of 0.6 Å/s.

*Solar Cell Characterization*

Photovoltaic performance was evaluated by measuring current density-voltage (*J-V*) curves under simulated AM 1.5G illumination (100 $mW/cm^2$). A MiniSol Sun Simulator (Model LSH-7320, Newport Corporation) was used as the light source, with the intensity calibrated using a standard silicon reference cell (Model 91150V, Newport Corporation). Data acquisition and voltage sweeps were performed using a Keithley 2400 SourceMeter Unit (SMU).

*Scanning Electron Microscopy*

Surface morphology was characterized using a Hitachi SU 8230 ultra-high resolution cold-field emission scanning electron microscope (FE-SEM).

*Time-Resolved Photoluminescence*

Spatially resolved PL lifetime mapping and decay kinetics were measured using a time-resolved fluorescence microscope (MicroTime 100, PicoQuant). The PL decay curves were fitted using a standard tri-exponential decay model to extract the carrier lifetimes:

$$I(t) = A_1\, exp\left(-\frac{t}{\tau_1}\right) + A_2\, exp\left(-\frac{t}{\tau_2}\right) + A_3\, exp\left(-\frac{t}{\tau_3}\right)$$

where $I(t)$ is the PL intensity at time $t$, $A_i$ represents the fractional amplitude of each decay component, and $\tau_i$ denotes the corresponding carrier lifetime. The average carrier lifetime ($\tau_{average}$) was calculated using the amplitude-weighted equation: $\tau_{average} = \sum A_i \tau_i^2 / \sum A_i \tau_i$ [47].

*Conductive AFM*

Conductive atomic force microscopy (cAFM) measurements were performed in an Asylum Research Cypher AFM (Oxford Instruments) using a Pt/Ir coated AFM tip (ElectriMulti75-G (Budget Sensors)) and a dual-gain ORCA tip holder. DN-DKL was implemented in the Cypher AFM via AEcroscopy platform (https://yongtaoliu.github.io/aecroscopy.pyae/welcome_intro.html) that was developed for AI/ML-driven autonomous microscopy experiments.